\documentclass[12pt,compress,nosort]{JHEP3}
\usepackage{exscale,latexsym}
\usepackage[dvips]{graphicx}
\usepackage{axodraw}
\def\Year{\expandafter\eatPrefix\the\year}
\newcount\hours \newcount\minutes
\def\monthname{\ifcase\month\or
January\or February\or March\or April\or May\or June\or July\or
August\or September\or October\or November\or December\fi}\def\shortmonthname{\ifcase\month\orx
Jan\or Feb\or Mar\or Apr\or May\or Jun\or Jul\or
Aug\or Sep\or Oct\or Nov\or Dec\fi}

\def\TimeStamp{\hours\the\time\divide\hours by60%
\minutes -\the\time\divide\minutes by60\multiply\minutes by60%
\advance\minutes by\the\time%
${\rm \shortmonthname}\cdot   \if\day<10{}0\fi\the\day\cdot   \the\year
\qquad\the\hours:\if\minutes<10{}0\fi\the\minutes$}

\newskip\humongous \humongous=0pt plus 1000pt minus 100pt
\def\caja{\mathsurround=0pt}
\def\eqalign#1{\,\vcenter{\openup1\jot \caja
       \ialign{\strut \hfil$\displaystyle{##}$&$
        \displaystyle{{}##}$\hfil\crcr#1\crcr}}\,}
\newif\ifdtup

\newcounter{eqnumber}[section]
\renewcommand{\theeqnumber}{\thesection.\arabic{eqnumber}}
\def\equn{\refstepcounter{eqnumber}
\eqno({\rm \theeqnumber})
}

\def\npb#1#2#3{{\rm Nucl. Phys. B}{\bf \ #1}, #3 (#2)}

\def\hepph#1{[hep-ph/#1]}

\def\tr{\mathop{\rm tr}\nolimits}

\newbox\charbox
\newbox\slabox
\def\s#1{{      % Feynman slash
        \setbox\charbox=\hbox{$#1$}
        \setbox\slabox=\hbox{$/$}
        \dimen\charbox=\ht\slabox
        \advance\dimen\charbox by -\dp\slabox
        \advance\dimen\charbox by -\ht\charbox
        \advance\dimen\charbox by \dp\charbox
        \divide\dimen\charbox by 2
        \raise-\dimen\charbox\hbox to \wd\charbox{\hss/\hss}
        \llap{$#1$}
}}

\def\spa#1.#2{\left\langle#1\,#2\right\rangle}
\def\spb#1.#2{\left[#1\,#2\right]}
\def\lor#1.#2{\left(#1\,#2\right)}

\catcode`@=11  % Make @ letter.

\def\Slash#1{\hskip 0.05 cm \slash\hskip -0.22 cm #1}

\def\eps{\epsilon}
\def\alp{\alpha}
\def\Ord{{\cal O}}

\def\e{\epsilon}

\def\la{\langle}
\def\ra{\rangle}

\def\lsl{\not{\hbox{\kern-2.3pt $\ell$}}}
\def\ksl{\not{\hbox{\kern-2.3pt $k$}}}

\def\rg{r_{\Gamma}}

\def\spa#1.#2{\left\langle#1\,#2\right\rangle}
\def\spb#1.#2{\left[#1\,#2\right]}
\def\lor#1.#2{\left(#1\,#2\right)}

\def\sand#1.#2.#3{%
  \left\langle\smash{#1}{\vphantom1}\right|{#2}%
  \left|\smash{#3}{\vphantom1}\right\rangle}
\def\sandp#1.#2.#3{%
  \left\langle\smash{#1}{\vphantom1}^{-}\right|{#2}%
  \left|\smash{#3}{\vphantom1}^{+}\right\rangle}
\def\sandpp#1.#2.#3{%
  \left\langle\smash{#1}{\vphantom1}^{+}\right|{#2}%
  \left|\smash{#3}{\vphantom1}^{+}\right\rangle}
\def\sandmm#1.#2.#3{%
  \left\langle\smash{#1}{\vphantom1}^{-}\right|{#2}%
  \left|\smash{#3}{\vphantom1}^{-}\right\rangle}
\def\sandpm#1.#2.#3{%
  \left\langle\smash{#1}{\vphantom1}^{+}\right|{#2}%
  \left|\smash{#3}{\vphantom1}^{-}\right\rangle}
\def\sandmp#1.#2.#3{%
  \left\langle\smash{#1}{\vphantom1}^{-}\right|{#2}%
  \left|\smash{#3}{\vphantom1}^{+}\right\rangle}

\def\Aloop{A^{\rm 1-loop}}

\def\Aloop{A^{\rm 1-loop}}

\def\Li{\mathop{\hbox{\rm Li}}\nolimits}

\def\Li{\mathop{\hbox{\rm Li}}\nolimits}

\def\tr{\mathop{\hbox{\rm tr}}\nolimits}

\def\L{\left(}\def\R{\right)}

\def\L{\left(}\def\R{\right)}

\def\BR#1#2{\la#1|{K_{abc}}|#2\ra}
\def\BR#1#2{[#1|{K}|#2\ra}

\def\NeqEight{{\cal N} = 8}
\def\NeqFour{{\cal N} = 4}
\def\NeqOne{{\cal N} = 1}

\title{Analytic Structure  of Three-Mass Triangle Coefficients}

\author{N.~E.~J.~Bjerrum-Bohr${}^{1}$, David~C.~Dunbar${}^2$
and Warren~B.~Perkins${}^2$
\\
\\
${}^1$Institute for Advanced Study,\\
Princeton, NJ 08540, USA\\
\\
${}^2$Department of Physics,\\
Swansea University,\\
Swansea, SA2 8PP, UK }

\abstract{
``Three-mass triangles'' are a class of integral functions appearing
in one-loop gauge theory amplitudes.  We discuss how the complex
analytic properties and singularity structures of these amplitudes can
be combined with generalised unitarity techniques to produce compact
expressions for three-mass triangle coefficients. We present formulae
for the $\NeqOne$ contributions to the $n$-point NMHV amplitude.  }


\begin{document}

\section{Introduction}

A general $n$-point one-loop amplitude in a massless theory such as
QCD can be expanded in terms of integral functions,
$$
 \Aloop_n=\sum_{i\in \cal C}\, c_i\, I_4^{i}
 +\sum_{j\in \cal D}\, d_{j}\, I_3^{j}
 +\sum_{k\in \cal E}\, e_{k} \,   I_2^{k}
+R\,,
\equn\label{basisequn}
$$
where $c_i,d_i,e_i$ and $R$ are rational functions and the $I_4$,
$I_3$, and $I_2$ are scalar box, triangle and bubble functions
respectively.  The mathematical form of these integral functions
depends on whether the momenta flowing into a vertex are null
(massless) or not (massive). This decomposition  suggests a ``divide and
conquer'' approach to evaluating one-loop amplitudes where different
techniques are used to evaluate the different types of coefficient.

In principle, traditional Feynman diagram techniques, combined with
reduction strategies can be used to compute the integral
coefficients~\cite{Feynman,Reduction}. Considerable progress has been
made in implementing this strategy, however the degree of complexity
rises rapidly with the number of external legs and the current state of
the art is in the computation of five and six-point
amplitudes (see for example~\cite{EGZ,StateOfFeynman}).

Alternate approaches based on the physical properties of amplitudes
have proved competitive or superior, particularly in computing
amplitudes with enhanced symmetry, such as those appearing in
supersymmetric theories, or for amplitudes with particular helicity
configurations.  Two particularly powerful methods have been those
based on unitarity and factorisation. The conjectured duality
between the perturbation theory of gauge theories and string
theory~\cite{Witten:2003nn} has provided added insight to these
approaches, particularly with respect to complex factorisation.

The unitarity method~\cite{BDDKa,BDDKb}, combined with a knowledge
of a basis set of integral functions for an amplitude, provides a
systematic way of calculating loop amplitudes. Two-particle cuts
provide sufficient information to identify many of the coefficients
in eqn.~(\ref{basisequn}) particularly in cases where the amplitude
is
``cut-constructible''~\cite{BDDKa,BDDKb,Bern:1997sc,BBCFsusyone,Britto:2006sj,OtherUnitarity,Binoth:2007ca,Britto:2006fc}.
In addition, extending to $D$-dimensional unitarity, in principle,
provides the information to calculate the rational
parts~\cite{Dunitarity}. Three and four particle cuts may also be
used to identify coefficients of triangle and box
functions~\cite{Bern:1997sc,BrittoUnitarity,Bidder:2005ri}. For the
box coefficients, $c_i$, quadruple cuts~\cite{BrittoUnitarity} are
particularly simple since the loop momentum integration is frozen by
the insertion of four $\delta$-functions.

The analytic structure of the cut integrals appearing in the
unitarity method can also be exploited to obtain coefficients. For
example, the ``holomorphic anomaly'' provided an insight into the
differing analytic properties of various integral functions in the
two-particle cuts~\cite{Cachazo:2004dr,Britto:2004nj,Bidder:2004tx}.
Various techniques have been developed to identify the integral
coefficients based on the analytic structure of the integrand of the
cut~\cite{BBCFsusyone,Britto:2006sj,Ossola}.

In this paper we explore some recent suggestions
for evaluating the coefficients of the ``three-mass triangle''
integral functions $I_3^{3m}(K_1,K_2,K_3)$
($K_i^2\neq 0$)
by using the analytic structure of the triple cut.
In ref.~\cite{Forde:2007mi} an algebraic technique was presented for
obtaining these coefficients.
In this
paper we review and refine this technique and present a version that uses
a single contour integration.

Although we can divide the amplitude into separate coefficients, in
general, different integral coefficients are related by a rich web
of ``spurious singularities''.  These are singularities that are not
present in the full amplitude but which appear in individual
coefficients. This structure is particularly rich for the three-mass
triangles. We explore and use this to obtain compact expressions for
these coefficients in the $\NeqOne$ contributions to six gluon
scattering. These provide alternate forms to those originally
calculated in~\cite{BBCFsusyone}.
We present formulae for the $n$-point ``next-to-MHV'' (NMHV) $\NeqOne$
contribution and describe how results may be obtained in the $N=0$ and
beyond-NMHV cases.

\section{Organisation of Amplitudes}

For one-loop amplitudes  with external gluons in QCD it is
convenient to decompose the contribution from gluons circulating in
the loop into pieces corresponding to complex scalars or
supersymmetric multiplets circulating in the loop,
$$
 A_{n}^{\rm 1-loop}
 =A_{n}^{\;\NeqFour}
 -4 A_{n}^{\;\NeqOne\;\rm chiral}+
 A_{n}^{\;\rm scalar}\,.
\equn
$$
The $A_{n}^{\;\NeqFour}$ component consists entirely of box
integrals.  The terms we look at in this paper are the
$A_{n}^{\;\NeqOne\;\rm chiral}$ contributions. These amplitudes may
contain any of the integral functions (but not rational terms).
Furthermore we consider the amplitude to be colour ordered and focus
our attention on the leading in colour component from which the full
amplitude can be obtained~\cite{BDDKa,Colour}.

The integrals appearing in the amplitude may be box, triangle or
bubble functions. We are interested in the contributions of
three-mass triangles.  The relevant integral function is defined by,
$$
I_3^{3m} = i \L4\pi\R^{2-\e} \,\int {d^{4-2\e}p\over \L2\pi\R^{4-2\e}}
\; { 1 \over p^2 \L p-K_1\R^2 \L p+K_2 \R^2}\; ,
\equn
$$
and can be written as~\cite{ThreeMassTriangle,BDKintegrals},
$$
I_{3}^{3  m}
=\ {i\over \sqrt{\Delta_3}}  \sum_{j=1}^3
  \left[ \Li_2\left(-\left({1+i\delta_j \over 1-i\delta_j}\right)\right)
       - \Li_2\left(-\left({1-i\delta_j \over 1+i\delta_j}\right)\right)
  \right]\ + \ \Ord(\e)\,,
\equn
$$
where,
$$
\eqalign{
\delta_1 & =
{ K_1^2-K_2^2-K_3^2
\over
\sqrt{\Delta_3}} \; , \cr
\delta_2 & = {K_2^2- K_1^2-K_3^2 \over
\sqrt{\Delta_3}} \; , \cr
\delta_3 & = { K_3^2- K_1^2-K_2^2 \over
\sqrt{\Delta_3}}\; ,
\cr}
\equn
$$
and
$$
\Delta_3\equiv -(K_1^2)^2 - (K_2^2)^2 - (K_3^2)^2
+ 2 K_1^2K_2^2
+2 K_1^2K_3^2
+2K_2^2K_3^2\; .
\equn
$$

The other integral functions we will encounter can be obtained in
many places e.g.~\cite{BDDKb}. The one-mass triangles depend only on
the momentum invariant of the massive leg, $K^2$,
$$
I_{3}^{1 m} (K^2)= {\rg\over\e^2} (-K^2)^{-1-\eps} \equiv G(K^2)
\; ,
\equn
$$
where $\rg\equiv {\Gamma(1+\eps)\Gamma^2(1-\eps) /  \Gamma(1-2\eps) }$.
The
two-mass triangle integral,
$$
I_{3}^{2  m}(K_1^2,K_2^2) = {\rg\over\e^2}
{(-K_1^2)^{-\eps}-(-K_2^2)^{-\eps}\over
(-K_1^2)-(-K_2^2) }\, ,
\equn
$$
can be expressed as one-mass triangle functions,
$$
I_{3}^{2  m}(K_1^2,K_2^2) =
{ 1 \over (-K_1^2)-(-K_2^2) }\
\left( G(K_1^2)- G (K_2^2)
\right)\,,
\equn
$$
and we can drop these functions from our  basis of integral
functions in favour of $G(K^2)$ functions. The box functions may be
found in many places, for example ref.~\cite{BDKintegrals,BDDKa}. We
need the form of one of these, namely the integral function where
two adjacent legs are massless; the so-called ``two-mass hard''
function. If $k_1$ and $k_2$ are the null legs,  defining   $S
\equiv (k_1+k_2)^2$ and $T=(k_2+K_3)^2$ we have,
$$\hspace{0.4cm}
\eqalign{
 I_{4}^{2{ m}h}
&=\ { -2 \rg  \over S T }
\biggl\{
 -{1\over\e^2} \Bigl[ (-S)^{-\e} + (-T)^{-\e}
              - (-K_3^2 )^{-\e} - (-K_4^2)^{-\e} \Bigr]
\cr &\hspace{-0.8cm}
   - {1\over2\e^2}
    { (-K_3^2)^{-\e}(-K_4^2))^{-\e}
     \over (-S)^{-\e} }
   + {1\over 2} \ln^2\left({ S \over T }\right)
   + \Li_2\left(1-{ K_3^2 \over T}\right)
   + \Li_2\left(1-{ K_4^2\over T }\right)
  \biggr\}  \, .
\cr}\hspace{-1cm}
\equn
$$

The coefficients of the integral functions will be expressed as
rational functions of spinor inner-products~\cite{SpinorHelicity},
$\spa{j}.{l} \equiv \langle k^-_j | k^+_l \rangle $, $\spb{j}.{l}
\equiv \langle k^+_j | k^-_l \rangle $, where $| k_i^{\pm}\ra $ is a
massless Weyl spinor with momentum $k_i$ and chirality $\pm$.  We
use notation where,
$$
\la k_a^+|\Slash{K}_{bcd}|k_e^+\ra \equiv [a|K_{bcd}|e\ra =
\spb{a}.b \spa{b}.e +
\spb{a}.c \spa{c}.e +
\spb{a}.d \spa{d}.e \; .
\equn
$$
As in twistor-space studies we define,
$$
\lambda_i \;=\; | k^+_i \rangle
\; , \;\;\;
\tilde\lambda_i \;= \;| k^-_i \rangle
\; .
\equn
$$

\section{Singularity Structure of Six-Point Three-Mass Triangles}

In this section we look at the three-mass triangle integrals found
in six-point one-loop gluon scattering amplitudes.  The only
non-vanishing three-mass triangle coefficients appear in the NMHV
amplitudes, of which there are three inequivalent forms:
$$
A(1^-,2^-,3^-,4^+,5^+,6^+)\,,
\;\;\;
A(1^-,2^-,3^+,4^-,5^+,6^+)\,,
\;\;\;
A(1^-,2^+,3^-,4^+,5^-,6^+)\,.
\equn
$$
The first of these was calculated in ref.~\cite{Bidder:2004tx} and
contains no three-mass triangles, as can be seen from the
triple-cuts. The remaining two were computed in
ref.~\cite{BBCFsusyone} using the analytic structure of the
two-particle cuts. Although correct (as verified by numerical
comparison to a Feynman diagram calculation~\cite{EGZ}), these
expressions contain irrational expressions involving the square root
of the Gram determinant of the three-mass triangle,
$\sqrt{\Delta_3}$.  We  produce expressions with the correct
singularity structure which explicitly do not contain these
irrational terms.

We start by considering the amplitude $A(1^-,2^+,3^-,4^+,5^-,6^+)$.
As for any supersymmetric amplitude, the cancellations occurring at
one-loop imply that no rational terms appear~\cite{BDDKb}. Further,
by examining the unitarity cuts, we see that only one-mass and
two-mass hard boxes appear and, as discussed above, we choose a
basis where the two-mass triangles are replaced by one-mass triangle
functions, $G(K^2)$. We thus have,
$$\hspace{-0.1cm}
\eqalign{ A^{\,\NeqOne}(1^-,2^+,  3^-,4^+,  5^-, 6^+)\;=\; & \sum_{i=1}^6
c_{i}^{1m}\;I_4^{1m\; i} + \sum_{i=1}^6 c_{i}^{2mh}\;I_4^{2mh\; i}
+\sum_{i=1}^6  d_i\;G(s_{i\,i+1}) \cr &\hspace{-3.5cm}+\sum_{i=1}^3
d'_i\;G(t_{i\,i+1\,i+2}) +d_1^{3m}\; I_3^{3m\; 1} + d_2^{3m}\; I_3^{3m\;
2} +\sum_{i=1}^6  e_i I_2^{2,i} +\sum_{i=1}^3 e_i' I_2^{3,i}\,, \cr}
\hspace{-2cm}\equn
$$
where both of the three-mass triangles,
$I_3^{3m\; 1}( K_{12},K_{34},K_{56} )$ and
$I_3^{3m\; 2}( K_{61},K_{23},K_{45} )$, appear.
The labelling of functions is specified below,

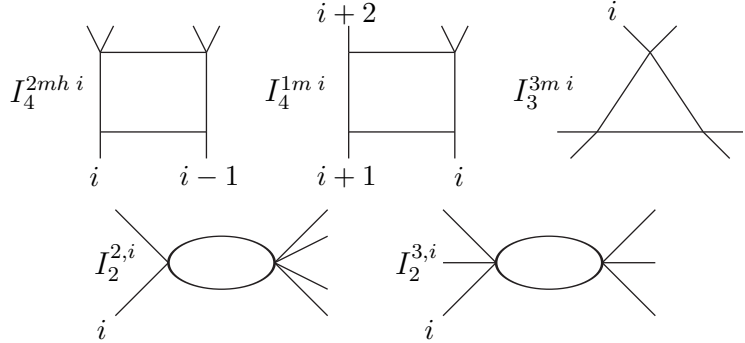
\begin{center}\hspace{1cm}
\begin{picture}(90,55)(0,0)
\Text(0,25)[c]{$I_4^{2mh\; i}$}
\Line(20,10)(60,10)
\Line(20,40)(60,40)
\Line(20,10)(20,40)
\Line(60,40)(60,10)
\Line(20,10)(20,0)
\Line(60,10)(60,0)
\Line(20,40)(15,50)
\Line(20,40)(25,50)
\Line(60,40)(55,50)
\Line(60,40)(65,50)
\Text(16,-7)[l]{\small $i$}
\Text(50,-7)[l]{\small $i-1$}
\end{picture}
\begin{picture}(90,55)(0,0)
\Text(0,25)[c]{$I_4^{1m\; i}$}
\Line(20,10)(60,10)
\Line(20,40)(60,40)
\Line(20,10)(20,40)
\Line(60,40)(60,10)
\Line(20,10)(20,0)
\Line(60,10)(60,0)
\Line(20,40)(20,50)
\Line(60,40)(55,50)
\Line(60,40)(65,50)
\Text(8,-7)[l]{\small $i+1$}
\Text(8,55)[l]{\small $i+2$}
\Text(48,50)[l]{\small $\null$}
\Text(57,50)[l]{\small $\null$}
\Text(68,50)[l]{\small $\null$}
\Text(60,-7)[l]{\small $i$}
\end{picture}
\begin{picture}(90,55)(0,0)
\Text(0,25)[c]{$I_3^{3m\; i}$}
\Text(24,56)[l]{\small $i$}
\Line(20,10)(40,40)
\Line(20,10)(60,10)
\Line(60,10)(40,40)
\Line(20,10)(5,10)
\Line(20,10)(10,0)
\Line(60,10)(75,10)
\Line(60,10)(70,0)
\Line(40,40)(30,50)
\Line(40,40)(50,50)
\end{picture}\vspace{0.3cm}
\begin{picture}(120,55)(0,0)
\Text(0,25)[c]{$I_2^{2,i}$}
\Text(-7,0)[l]{\small $i$}
\Oval(40,25)(10,20)(0)
\Line(60,25)(80,45)
\Line(60,25)(80,5)
\Line(60,25)(80,35)
\Line(60,25)(80,15)
\Line(20,25)(0,5)
\Line(20,25)(0,45)
\end{picture}
\begin{picture}(70,55)(0,0)
\Text(-10,25)[c]{$I_2^{3,i}$}
\Text(-7,0)[l]{\small $i$}
\Oval(40,25)(10,20)(0)
\Line(60,25)(80,45)
\Line(60,25)(80,5)
\Line(60,25)(80,25)
\Line(20,25)(0,5)
\Line(20,25)(0,45)
\Line(20,25)(0,25)
\end{picture}
\end{center}

We shall see how much of the amplitude can be reconstructed from the
singularity structure: both real and spurious.  Our starting point
is the box coefficients~\cite{Bidder:2004vx},
$$\hspace{1cm}
\eqalign{
c_1^{1m} & =
\displaystyle  i{ \BR{2}{5}^2 \BR15 \BR35 \over \spb1.3^2 \spa4.5 \spa5.6 \BR14\BR36 K^2  }
\times { -s_{12}s_{23} \over 2 }
\,,\hskip 0.5 truecm
K=K_{123}\,,
\cr
c_{4}^{2mh} &=
\displaystyle i{ \BR{2}{5}^2  \BR35 \BR24
\over \spb1.2 \spa5.6 \BR36 \BR14 \BR34^2   }\
\times { -s_{34} K^2  \over 2}
\,,\hskip 0.5 truecm
K=K_{123}\,.
\cr}
\equn
$$
We will see that these completely determine the coefficients of the
one-mass triangles and determine much of the three-mass triangle
coefficients.

\subsection{Infra-Red Singularities}

One of the major constraints on the triangle coefficients comes from
requiring that the amplitude has the correct infra-red
singularities. The box integral functions with massless legs and the
one-mass triangle functions both contain $\ln(K^2)/\eps$
singularities.  We then have,
$$
\left[
d_i\; G(K^2) +\sum c_i\; I_4^{i} \right]_{\ln (K^2)/\eps}
= 0
\; ,
\equn
$$
for the $\NeqOne$ chiral multiple where such singularities do not
appear in the amplitude. These constraints fix the coefficients of
the $G(K^2)$ in terms of the box coefficients.

For the $\NeqOne$ multiplet, one could choose a basis in which the
coefficients of the one-mass triangles are zero. There are several
options for doing so: firstly one can choose the basis of ``six
dimensional scalar box functions'' as in~\cite{Bidder:2004tx} or one
can choose a basis of functions where the IR singularities have been
subtracted as in~\cite{BBCFsusyone}.  In the first case the $D=4$
boxes and $D=6$ boxes are related, using the notation of
ref.~\cite{BDKintegrals},
$$
I_4^{D=4}  = { 1 \over 2 N_4 }
\Biggl[
\sum_i \alpha_i \gamma_i I_3^{(i)}  +(-1+2\eps)\hat \Delta_4  I_4^{D=6}
\Biggr]
\; ,
\equn\label{BoxReduction}
$$
where $I_3^{(i)}$ is the descendant triangle in which the $i$-th
propagator is deleted. If we change the basis of box integral
functions, the coefficients of the triangles, including the
three-mass, are shifted,
$$
d_i\; \longrightarrow\; d_i + { \alpha_i \gamma_i \over 2 N_4 } c_{box}\,.
\equn
$$

Although transforming to this basis is instructive, we will continue
to use the basis of $D=4$ integral functions.

\subsection{Spurious Singularities}
Looking at the box coefficients we see that factors such as,
$$
{1 \over [3 | K_{123} | 4 \ra^2  }
\;\;\;\hbox{\rm and }
\;
{1 \over \spb1.3^2 }\,,
\equn
$$
appear. The first of these is singular when the momenta are arranged
such that,
$$
k_1^\mu+k_2^\mu \;=\; \alpha k_3^\mu +\beta k_4^\mu
\; ,
\equn
$$
and such singularities are termed co-planar\footnote{For real
momenta $[4|K|3\ra$ and $[3|K|4\ra$ vanish simultaneously. However,
by continuing to complex momenta we can find a point where only one
of them vanishes, e.g. if $k_1+k_2\sim
\lambda_x\bar\lambda_3+\lambda_4\bar\lambda_y$, $[3|K|4\ra=0$ but
$[4|K|3\ra\neq 0$.}. These singularities are {\it spurious}, meaning
they may appear in individual terms within an amplitude but
disappear when the entire amplitude is constructed. These coplanar
singularities do not cancel amongst the boxes, but cancel between
the boxes and the other integral functions.  For real momenta these
singularities occur when,
$$
t_{123}^2 +(s_{34}-s_{56}-s_{12} )\;t_{123} +s_{56}\;s_{12} \;=\;0\,.
\equn
$$

At the coplanar singularity the link between the box coefficients
and  the one-mass triangles implies that the latter also have
coefficients with quadratic singularities. However the cancellation
of the spurious singularities extends beyond these functions. At
this singularity the dilogarithms within the two mass hard function
simplify since $s_{12}=\alpha\beta s_{34}$,
$s_{56}=(1+\alpha)(1+\beta)s_{34}$ and
$t_{123}=\alpha(1+\beta)s_{34}$, leading to,
$$
I^{2m h}_4
\sim \left(   \Li_2\left( 1- { \alpha\over (1+\alpha) } \right) +\Li_2\left( 1-{1+\beta\over \beta} \right) \right)
\; .
\equn
$$
These must cancel against another integral function containing
dilogarithms: the three-mass triangle being the only possibility.
The cancellation of spurious singularities can thus be expressed as,
$$
\left[
c_i I_4^{i} + d_i I_3^{1m\; i} +d^{3m} I_3^{3m} \right]_{[a|K|b\ra=0}
\;=\; {\rm finite}\,.
\equn
$$
This imposes a significant constraint on the three-mass triangle
coefficient. One approach is to change basis to one where this
cancellation is automatic.  This process is essentially the same as
that of ref.~\cite{Bern:1997sc,Campbell:1996zw} where the three-mass triangle
functions arise in $e^+ +e^- \longrightarrow {\rm four\; parton}$
scattering. We can generate this combination using the identity
(\ref{BoxReduction}),
$$
I_4^{D=4}  - { 1 \over 2 N_4 }
\Biggl[
\sum_i \alpha_i \gamma_i I_3^{(i)}
\Biggr] =(-1+2\eps){ \hat \Delta_4 \over 2 N_4 } I_4^{D=6}
\; .
\equn
$$
For the two-mass hard box we have,
$$
{ \hat \Delta_4 \over 2 N_4 }
=
-2\left(
{
\tr ( \Slash{k_{i-1}}  \Slash{P}\Slash{k_{i}}\Slash{P} )
\over S T^2 }
\right)
=-2 {[ i | P | i-1\ra [ i-1 | P | i \ra
\over s_{i-1i} (P^2)^2 }
\; .
\equn
$$
Thus $\hat \Delta_4/2 N_4 \longrightarrow 0$ at the coplanar
singularity and since $I_4^{D=6}$ is finite at this (unphysical)
singularity we must have,
$$
I_4^{D=4}  - { 1 \over 2 N_4 }
\Biggl[
\sum_i \alpha_i \gamma_i I_3^{(i)}
\Biggr]
 \longrightarrow 0\,,
\equn
$$ at the coplanar singularity. Up to a scaling factor this
is precisely the $Ls_1$ function of ref.~\cite{Bern:1997sc}. As this
combination includes a three-mass triangle, the coefficient of this
three-mass triangle in the $D=4$ basis must contain a term,
$$
-{\alpha_i \gamma_i \over 2 N_4 } c_{4}^{2m h}
\; ,
\equn
$$
which suggests the term,
\def\Spa(#1,#2){\left\langle#1\,#2\right\rangle}
\def\Spb(#1,#2){\left[#1\,#2\right]}

\def\Spba(#1,\{#2,#3,#4\},#5){\left [#1|K_{#2#3#4}|#5\right \rangle}

\def\Spaa(#1,\{#2,#3\},\{#4,#5\},#6){\left\langle#1|K_{#2#3}K_{#4#5}|#6\right\rangle}

\def\Spab(#1,\{#2,#3,#4\},#5){\left [ #5|K_{#2#3#4}|#1\right \rangle}

$$
\frac{-\Spab(4,\{1,2,3\},2) \Spab(5,\{1,2,3\},2)
\Spab(5,\{1,2,3\},3)}{\Spab(4,\{1,2,3\},1) \Spab(4,\{1,2,3\},3)^2
\Spab(6,\{1,2,3\},3) t_{123}} \frac{ \Spab(5,\{1,2,3\},2) (2 s_{12}
s_{56} -(s_{12}+s_{56}-s_{34})\; t_{123})}{2 \Spa(5,6) \Spb(1,2)
 }\,,
\equn
$$
within $d^{3m}$.  Since the three-mass triangle is the ``daughter''
of three different two-mass-hard boxes,  each with a different
quadratic singularity, we have three such terms in the coefficient.

This expression gives an amplitude from which the quadratic spurious
singularity is absent. However we have introduced a further
fictitious singularity : a $t_{123}$ pole.  The three-mass triangle
should not contain such a pole. We can ``fix''  this by adding an
extra term, giving,
$$
\hspace{-0.3cm}
\frac{-\Spab(4,\{1,2,3\},2) \Spab(5,\{1,2,3\},2) \Spab(5,\{1,2,3\},3)}{\Spab(4,\{1,2,3\},1)\! \Spab(4,\{1,2,3\},3)^2\! \Spab(6,\{1,2,3\},3) \!t_{123}}
\!\!
\left(\!\frac{ \Spab(5,\{1,2,3\},2)\!
(2 s_{12} s_{56}\! -\!(s_{12}\!+\!s_{56}\!-\!s_{34}) t_{123})}{2 \Spa(5,6) \Spb(1,2) \Spab(4,\{1,2,3\},3)
 }
\!+\!\Spa(1,3)\! \Spb(6,4)\!  \right)\!.
\equn
$$
This process fixes the leading quadratic coplanar pole. Fixing the
remaining linear singularity
 gives more terms in the amplitude.
Repeating the process as before we can deduce that,
$$
{1 \over [3 | K | 4 \ra^2 }
\left(
I_4^{D=4}  - { 1 \over 2 N_4 }
\Biggl[
\sum_i \alpha_i \gamma_i I_3^{(i)}
\Biggr]
+{ \Delta_4 \over (2N_4)^2 }
\Biggl[
\sum_i \alpha_i \gamma_i I_3^{D=6,(i)}
\Biggr]
\right)
\longrightarrow {\rm finite}\,,
\equn
$$
at the coplanar singularity.
The function,
$$
J_4 \equiv
\left(
I_4^{D=4}  - { 1 \over 2 N_4 }
\Biggl[
\sum_i \alpha_i \gamma_i I_3^{(i)}
\Biggr]
+{ \Delta_4 \over (2N_4)^2 }
\Biggl[
\sum_i \alpha_i \gamma_i I_3^{D=6,(i)}
\Biggr]
\right)\,,
\equn
$$
is a combination of the $D=4$ two-mass box, $D=4$ triangle functions
and $D=4$ bubble functions. If we took the box coefficients and used
the $J_4$ functions as a basis rather than the $I_4$ functions, we
could extend the box contributions,
$$
\sum_i c_i I_4^i \longrightarrow \sum_i c_i J_4^i\,,
\equn
$$
to obtain an expression containing much of the three-mass triangle
and bubble contributions to the amplitude. This would be an
expression without $[3|K_{123}|4\ra$, $[1|K_{561}|2\ra$ or
$[5|K_{345}|6\ra$ singularities. It may however contain linear
singularities due to $[5|K_{561}|2\ra$, $[1|K_{456}|4\ra$ or
$[3|K_{123}|6\ra$ vanishing.

\def\wbpnorm{\times (-i)}
Looking carefully at the full singularity structure, after some
trial and error,  we are led to the expression for the three-mass
triangle coefficient,
$$\hspace{-0.3cm}
\eqalign{ &d_{3m}^{[ \{ 1^-2^+\} ,\{ 3^-4^+ \}, \{ 5^-6^+\}
]}\wbpnorm = \cr &\hskip -10pt -\frac{\Spab(1,\{3,4,5\},4)\!
\Spab(1,\{3,4,5\},5)\! \Spab(6,\{3,4,5\},4)}{\Spab(2,\{3,4,5\},5)\!
\Spab(6,\{3,4,5\},3)\! \Spab(6,\{3,4,5\},5)\! t_{345}} \hskip
-4pt\left( \frac{ \Spab(1,\{3,4,5\},4) (2 s_{12} s_{34}\! +\!(
s_{56}\! -\!s_{12}\!-\!s_{34}) t_{345})} {2 \Spa(1,2) \Spb(3,4)
\Spab(6,\{3,4,5\},5)} \!+\!\Spa(3,5)\! \Spb(2,6)\right) \cr & \hskip
-10pt - \frac{ \Spab(2,\{5,6,1\},6)\! \Spab(3,\{5,6,1\},1)\!
\Spab(3,\{5,6,1\},6) }{\Spab(2,\{5,6,1\},1)\! \Spab(2,\{5,6,1\},5)\!
\Spab(4,\{5,6,1\},1) \!t_{561}}
\hskip -4pt\left(
\frac{ \Spab(3,\{5,6,1\},6) (2 s_{34} s_{56}\!+\!( s_{12}\! -\!s_{34}\!-\!s_{56}) t_{561})}{2 \Spa(3,4) \Spb(5,6) \Spab(2,\{5,6,1\},1)
}
\!+\!\Spa(5,1)\! \Spb(4,2)
\right)
\cr &
\hskip -10pt-
\frac{\Spab(4,\{1,2,3\},2)\! \Spab(5,\{1,2,3\},2)\! \Spab(5,\{1,2,3\},3)}{\Spab(4,\{1,2,3\},1)\! \Spab(4,\{1,2,3\},3)\! \Spab(6,\{1,2,3\},3) \!t_{123}}
\hskip -4pt\left(\frac{\Spab(5,\{1,2,3\},2)
(2 s_{12} s_{56}\! -\!(s_{12}\!+\!s_{56}\!-\!s_{34}) t_{123})}{2 \Spa(5,6) \Spb(1,2)
\Spab(4,\{1,2,3\},3)  }
\!+\!\Spa(1,3)\! \Spb(6,4)  \right)
\cr &
\hskip -20pt-\biggl(
\frac{\Spa(1,3) \Spa(3,5) \Spb(2,6) \Spb(3,4)+\Spa(1,3) \Spa(1,5) \Spb(1,2) \Spb(4,6)+\Spa(1,5) \Spa(3,5) \Spb(2,4) \Spb(5,6)}
{\Delta_3}
\biggr)\times
\cr &
\hspace{-.6cm}\Biggl(\!\!
\frac{ \Spab(1,\{\!3,4,5\!\},5)\! \Spab(6,\{\!3,4,5\!\},4)\! ( t_{\!345}\!-\!t_{\!346\!} )}{\Spab(2,\{\!3,4,5\!\},5)\!
\Spab(6,\{\!3,4,5\!\},3)\! \Spab(6,\{\!3,4,5\!\},5)}
\!+\!\frac{ \Spab(2,\{\!5,6,1\!\},6)
\!\Spab(3,\{\!5,6,1\!\},1)\! ( t_{\!561}\!-\!t_{\!562\!} )}
{\Spab(2,\{\!5,6,1\!\},1)\! \Spab(2,\{\!5,6,1\!\},5)\! \Spab(4,\{\!5,6,1\!\},1)}
\!+\!\frac{ \Spab(4,\{\!1,2,3\!\},2)\! \Spab(5,\{\!1,2,3\!\},3)\! (t_{\!123}\!-\!t_{\!124\!})}
{\Spab(4,\{\!1,2,3\!\},1)\! \Spab(4,\{\!1,2,3\!\},3)\! \Spab(6,\{\!1,2,3\!\},3)}
\cr &
\hskip -40pt
\quad\quad\quad
\quad\quad\quad
\hspace{2.3cm}
-2 \frac{  \Spab(2,\{2,3,4\},6)
\Spab(4,\{4,5,6\},2) \Spab(6,\{6,1,2\},4)
-\Spab(1,\{3,4,5\},5)\! \Spab(3,\{5,6,1\},1)
\Spab(5,\{1,2,3\},3)}{\Spab(2,\{5,6,1\},5)
\Spab(4,\{4,5,6\},1)
\Spab(6,\{4,5,6\},3)}\!
\Biggr).
\cr}\equn
$$
We have confirmed this expression by comparison with a numerical
evaluation of the triple cut. This provides an alternative form for
the coefficient previously obtained in ref.~\cite{BBCFsusyone}. Our
form is free of irrational expressions and has a more manifest
singularity structure.

We also obtain a rational form for the other six-point three-mass
triangle coefficient,
$$\hspace{-0.3cm}
\eqalign{
&d_{3m}^{[  \{2^- 3^+\}, \{4^- 5^+\},  \{6^+ 1^-\}  ]}\wbpnorm=
\cr
&
\hspace{0.6cm}{ \spb2.6   [2|K_{345}|4\ra  [6|K_{345}|4\ra     \over
\spb1.2    [2|K_{345}|3\ra  [2|K_{345}|5\ra   t_{345}                }
\!\Biggl(\!
{[6|K_{345}|4\ra ( 2 s_{61}s_{45} +(s_{23}-s_{61}-s_{45} )t_{345}  )\over
 2 \spb6.1\spa4.5  [2|K_{345}|3\ra }
+\spa1.2 \spb3.5
\!\Biggr)\!
\cr
&
+{\spa5.1 [3|K_{234}|5\ra [3|K_{234}|1\ra\over \spa5.6  [4|K_{234}|5\ra  [2|K_{234}|5\ra t_{234}}
\!\Biggl(\!
 { [3|K_{234}|1\ra (  2 s_{23}s_{61} +(s_{45}-s_{23}-s_{61})t_{234} )
\over
 2 \spb2.3\spa1.6 [4|K_{234}|5\ra   }
+\spa2.4  \spb6.5
\!\Biggr)\!
\cr
&
\hspace{1cm}+{   \spb1.3 \spa6.4  [3|K_{123}|4\ra\over
 \spb1.2 \spa5.6     [1|K_{123}|6\ra t_{123}              }
\!\Biggl(\!
{[3|K_{123}|4\ra ( 2 s_{45}s_{23} +(s_{61}-s_{45}-s_{23})t_{123} )\over
  2 \spb2.3  \spa4.5 [1|K_{123}|6\ra }
+\spa1.2\spb6.5
\!\Biggr)\!
\cr &
-\biggl({\spa4.2\spa2.1\spb3.2\spb6.5 +\spa4.1\spa2.1\spb6.1\spb3.5 +\spa4.2\spa4.1\spb4.5\spb3.6
   \over
\Delta_3
}
\biggr)
\cr &
\qquad\times\Biggl(
2{      \spb2.6 \spa6.5 \spb3.6 \spa6.4 - \spa5.1 \spb1.2 \spa4.1 \spb1.3
   \over
        \spb1.2\spa5.6 [2|K_{561}|5\ra                                     }
+{     \spb1.3  \spa4.6 (t_{123}-t_{623})
\over
       \spa5.6 \spb1.2  [1|K_{123}|6\ra  }
\cr &
\qquad\qquad\hspace{3.2cm} +{   \spb2.6   [2|K_{345}|4\ra ( t_{345}-t_{245})
\over
     \spb1.2 [2|K_{345}|5\ra  [2|K_{345}|3\ra  }
+{     \spa5.1   [3|K_{234}|5\ra(t_{234}-t_{235})
\over
       \spa5.6 [2|K_{234}|5\ra  [4|K_{234}|5\ra}
\Biggr)\,,
\cr}
\equn
$$
which we have again confirmed by comparison with a numerical
evaluation of the triple cut.

\section{Analytic evaluation of the Three-Mass Triangle Coefficients}

In this section we explore and refine some recent suggestions for
using the analytic structure of triple
cuts~\cite{Forde:2007mi,Ossola,Mastrolia:2006ki} to evaluate the
three-mass triangle coefficients.

Consider a triple cut in an amplitude where all three corners
are massive, 
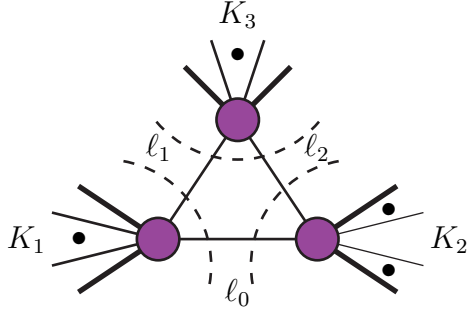
\vspace{0.6cm}
\begin{center}
\begin{picture}(40,60)(-20,50)
\Line(30,30)(70,40)
\Line(30,30)(70,20)
\SetWidth{2}
\Line(30,30)(60,50)
\Line(30,30)(60,10)
\SetWidth{1}
\Line(30,30)(-30,30)
\Line(-30,30)(0,75)
\Line(30,30)(0,75)
\Line(-30,30)(-70,40)
\Line(-30,30)(-70,20)
\SetWidth{2}
\Line(-30,30)(-60,50)
\Line(-30,30)(-60,10)
\SetWidth{1}
\Text(-60,30)[]{$\bullet$}
\Line(0,75)(-10,105)
\Line(0,75)(10,105)
\SetWidth{2}
\Line(0,75)(-20,95)
\Line(0,75)(20,95)
\Text(0,100)[]{$\bullet$}
\Text(57,41)[]{$\bullet$}
\Text(57,18)[]{$\bullet$}
 \SetWidth{1}
\DashCArc(45,20)(40,100,190){4}
\DashCArc(-50,20)(40,-10,80){4}
\DashCArc(00,100)(40,220,320){4}
\CCirc(30,30){8}{Black}{Purple}
\CCirc(-30,30){8}{Black}{Purple}
\CCirc(0,75){8}{Black}{Purple}
\Text(0,10)[]{$\ell_0$}
\Text(-30,65)[]{$\ell_1$}
\Text(30,65)[]{$\ell_2$}
\Text(80,30)[]{$K_2$}
\Text(-80,30)[]{$K_1$}
\Text(0,115)[]{$K_3$}
\end{picture}
\end{center}
\vspace{1.2cm}
$$
\eqalign{
C_3\,=\,
 \sum_{h_i \in {\cal S}'} & \int d^4\ell_1  \delta(\ell_0^2)
 \delta(\ell_1^2) \delta(\ell_2^2)
 A_1\Big((\ell_0)^{h_1}, i_m, \cdots, i_j, (-\ell_1)^{-h_2} \Big)
\cr
&
\hspace{-1.3cm}\,\times\,
A_2\Big((\ell_1)^{h_2}, i_{j+1} ,\cdots, i_l, (-\ell_2)^{-h_3} \Big)
\,\times\,
A_3\Big((\ell_2)^{h_3}, i_{l+1} ,\cdots, i_{m-1},  (-\ell_0)^{-h_1} \Big)\,,
\cr}
\hspace{-1cm}
\equn\label{cuteq}
$$ where the summation is over all possible intermediate states.  As
the momentum invariants, $K_m=k_{i_m}+k_{i_m+1}+\cdots +k_{i_j}$
etc, are all non-null, there exist kinematic regimes is which the
integration has non-vanishing support for real loop momentum. If we
expand the amplitude in terms of a basis of integral
functions~(\ref{basisequn}), the only integral functions
contributing to the triple cut are box functions and the specific
three-mass triangle for the cut,
$$
\eqalign{
C_3 &\;=\; \sum_i c_i
(I_4^i)_{\rm triple-cut} \;+\; d_{3m}( I_3^{3m} )_{\rm triple-cut}
\cr
&\;=\; \sum_i { c_i \over \kappa_i} \;+\;  d_{3m} { \pi \over 2 \sqrt{-\Delta_3} }\,,
\cr}
\equn
$$
For the two-mass hard and three-mass boxes that arise in the six and
seven-point examples we discuss, the cuts of the box integral
functions are,
$$
{1 \over \kappa^{ 2mh}}\;=\;\pm{\pi\over 2 (k_1+k_2)^2(k_2+K_3)^2
}\,,\;\; {1 \over \kappa^{ 3m}}\;=\;\pm{\pi\over 2\Bigl(
(k_1+K_2)^2(K_2+K_3)^2-K^2_2 K^2_4 \Bigr)}\, . \equn$$ We will
discuss the overall sign below.

Alternatively we can perform the cut integral~(\ref{cuteq}). The
triple cut is a one-parameter integral which can be calculated
using algebraic  methods~\cite{Forde:2007mi}. We  review the
procedure for the general triple cut emphasising the geometric
interpretation in the three-mass case as a contour integral.

The first step is to find a suitable parameterization of the cut
momenta which satisfy $l_i^2=0$ with $l_1=l_0-K_1$ and
$l_2=l_0+K_2$. As $\sum K_i=0$, the momenta $K_i$ define a plane.
Within this plane there exists a momentum, $a_0^\mu$, satisfying
$a_0^2=(a_0-K_1)^2=(a_0+K_2)^2$. Explicitly~\cite{Ossola},
$$
a_0^\mu = {K_2^2\over 2}{K_1\cdot K_2+K_1^2\over K_1^2K_2^2-(K_1 \cdot K_2)^2} K_1^\mu
          -{K_1^2\over 2}{K_1\cdot K_2+K_2^2\over K_1^2K_2^2-(K_1\cdot K_2)^2}K_2^\mu\,.
\equn
$$
In the three-mass case, $\vert a_0\vert\neq 0$, the cut momenta are
real and for $a_0$ time-like can be parameterised in the form,
$$
\ell_0^{\mu}\,=\, a_0^{\mu}+\rho(\cos\theta m^\mu + \sin\theta n^\mu) \,,\equn
$$
where $\rho=\sqrt{-a_0^2}$ and $0 \leq \theta \leq 2\pi$.  The
vectors  $m$ and $n$ are mutually orthogonal unit vectors which are
orthogonal to the $(K_1,K_2)$ plane; $(m \cdot n)\, =\,(m \cdot K_i)\, =\,(n
\cdot K_i)\, =\,0$. For $a_0$ space-like, a hyperbolic parameterization
can be used. If we now define the complex null momenta,
$r={\rho\over 2}(m+in)$ and $\overline r={\rho\over 2}(m-in)$, we
recover the parameterization used in
~\cite{Ossola,Forde:2007mi}\footnote{The $t$ in our parameterization
and that in~\cite{Forde:2007mi} are related by a scaling},
$$
\ell_i^{\mu}\,=\,t\, \overline r^{\mu}\,+\,\frac{1}{t}\, r^{\mu}+a_i^{\mu}\,,
\equn\label{ParaA}
$$
where $t={\rm e}^{i\theta}$, $a_1=a_0-K_1$ and $a_2=a_0+K_2$.

We can define null momenta $\hat K_i$ in the plane of the $K_i$ via~\cite{Forde:2007mi},
$$
\eqalign{
\hat K_1 =&  {\gamma^2 \over \gamma^2-K_1^2K_2^2 }\left(  K_1 -{K_1^2\over \gamma}  K_2 \right)\,,
\cr
\hat K_2 =&  {\gamma^2 \over \gamma^2-K_1^2K_2^2 }\left(  K_2 -{K_2^2\over \gamma}  K_1 \right)\,,
\cr}
\equn
$$
where $\gamma =K_1\cdot K_2 + {1\over 2} \sqrt{-\Delta_3}$.
In terms of the $\hat K_i$,
$$
r \sim \lambda_{K_2} \bar \lambda_{K_1} \;,\;\;
\bar r \sim \lambda_{K_1} \bar \lambda_{K_2} \;.\;\;
\equn
$$
(We will drop the ``hat'' on $K_i$ when it is clear from context that we are referring to the null form.)
%%% NEW

For the spinors
this parameterisation corresponds to,
$$
\lambda_{l_i}= t \lambda_{K_1} +\alp_{01} \lambda_{K_2}\,,
\;
\;
\bar\lambda_{l_i}= {1 \over t} \left( t\bar\lambda_{K_2}+\alp_{02} \bar\lambda_{K_1} \right)\,,
\equn\label{ParaEqn}
$$
where,
$$
\alpha_{01}= { K_1^2(\gamma -K_2^2) \over \gamma^2-K_1^2K_2^2  }\,,
\;\;\;
\alpha_{02}= { K_2^2(\gamma -K_1^2) \over \gamma^2-K_1^2K_2^2  }\,.
\equn
$$

With the parameterization (\ref{ParaA}) it is clear that the cut integration
becomes a contour integration over the complex variable $t$ with the
contour being the unit circle. The integral then becomes,
$$
\int d^4l  \prod_i \delta(\ell_i^2)(\bullet)\,  \longrightarrow \, \int dt J_t (\bullet)\,,
\equn
$$
where $J_t = 1/( 4 t \sqrt{\Delta_3} )$ is the Jacobian. 
Regarding $t$ as complex allows the integral to
be performed analytically using contour methods. In the three-mass
case the contour is well specified.

Parameterising the loop momenta according to (\ref{ParaA}) the
product of tree amplitudes $A_1\;A_2\;A_3$ is a rational function of
$t$. This rational function will have simple poles at $t=t_i\neq 0$
and, possibly, non-simple poles at $t=0$. Poles in this product at
$t=t_i\neq 0$ arise when one of the tree amplitudes factorises and
some momentum, $\hat P(t)$, becomes null:
$$
A\,\, \longrightarrow \hskip -1.0truecm
 {\atop _{\hat P^2\to 0}}
\,\,\hat A_L \,\,{1\over \hat P(t)^2}\,\, \hat A_R,
\equn$$
where $\hat A_L$ and $\hat A_R$ are tree amplitudes evaluated at the
momenta where $\hat P^2$ vanishes. In general $\hat P^2=0$ gives two
poles. For the six and seven-point examples we discuss, one of these
poles gives the box contribution while the other gives no
contribution. In these examples each box has at least  one massless
corner and we have  $\hat P=l\pm a$, where $a$ is the external
momentum of a massless corner. Poles arise when either
$\spa{l}.{a}=0$ or $\spb{l}.{a}=0$. The original tree amplitudes
will only contain one of these poles, so only one of the poles can
contribute to the triple cut. If the appropriate pole is inside the
contour of integration, the contribution to the triple cut is of the
form,
$$
2\pi i \,{\rm Res}\biggl({A_1A_2\hat A_{3L}\hat A_{3R} \over 4 t\sqrt{\Delta_3}\hat P^2}\biggr)\biggr\vert_{t=t_i}.
\equn$$
By comparison with the quadruple cut procedure, we see that the
product of on-shell tree amplitudes reproduces the box coefficient
up to a factor of 2. It is useful to compare the rest of this
expression to the triple cut of the corresponding scalar box,
$$\eqalign{
\int  dt J_t\,  { 1\over (l_0-P)^2}\,
=\int  {dt \over4  t\sqrt{\Delta_3}}\,  { 1\over \hat P^2 }\,.
\cr}\equn
$$
This has poles in identical positions, but both could in principle
contribute. Denoting the two $t$-values for which $(l_0-P)^2$
vanishes by $t_{\pm}$, we have,
$$\hspace{-0.6cm}
\eqalign{
{1 \over t(\ell_0-P)^2 }\, =\, {  -1 \over (2\overline r\cdot P)(t-t_+)(t-t_-) }
\,=\,{-1 \over (2\overline r\cdot P)(t_+-t_-)}
\left( { 1 \over t-t_+} -{ 1 \over t-t_-}
\right)\,.
\cr}\equn
$$
The two poles thus have equal but opposite residues.
In the three-mass case, $t_+$ and $t_-$ are the roots the quadratic equation,
$$
2\overline r\cdot P t^2 +(2a_0\cdot P-P^2)t+ 2r\cdot P=0\,,
\equn$$
so the product of the roots is,
$$
t_+t_-={r\cdot P \over \overline r\cdot P} \;\to\; \vert t_+t_-
\vert =1\,. \equn$$
One pole is always inside the unit circle and
one is outside. Thus the triple cut of the box function always gives
a contribution, but the sign depends on the kinematic point. In
contrast, the original triple cut integral only receives
contributions if the appropriate pole is inside the contour of
integration.

In general $(A_1\, A_2\, A_3)/ t$ can also have a pole at $t=0$ and
we denote the residue of this  pole by $\rho_0$. Using both
approaches to evaluate the triple cut integral we then have,
$$\eqalign{
C_3\,=\,
\sum_{i} 2\Theta(1-\vert t_i\vert){c_i\over\tau_i}  \, +\, { 2\pi \rho_0 \over \sqrt{-\Delta_3} }
\,=\,\sum_{i} -(-1)^{\Theta(1-\vert t_i\vert)}{ c_{i} \over \tau_i} \,+\,  d_{3m} { \pi \over 2 \sqrt{-\Delta_3} }\,,
\cr}\equn
$$
where, $\tau_i=-(-1)^{\Theta(1-\vert t_i\vert)}\kappa_i$.
We can rearrange this to give an expression free from $\Theta$ functions,
$$\eqalign{
{ 2\pi  \rho_0 \over \sqrt{-\Delta_3} }
\,=\,\sum_{i} -{ c_{i} \over \tau_i} \,+\,  d_{3m} { \pi \over 2 \sqrt{-\Delta_3} } \,\equiv\,{\cal S} \equiv
{\cal S}^{\rm box} +{\cal S}^{\rm triangle}  \,,
\cr}\equn
$$
which relates $\rho_0$ to a specific sum of box and triangle
contributions. The box contributions are readily calculated either
by quadruple cuts or by using the fact that they are half of the
$t\neq 0$ residues. The latter approach provides a realisation of
the quadruple cut procedure that is amenable to
automation~\cite{Forde:2007mi}.  The three-mass triangle
contribution to $\rho_0$ can thus readily be identified. A slightly
different formulation involving integration over two different
regions (corresponding to the interior and exterior of the unit
circle in this case) was presented in~\cite{Forde:2007mi}.

For any $n$-point NMHV amplitude the three tree amplitudes in the
triple-cut of a three-mass triangle are of MHV type. When we
parameterise the integral by $t$, each tree amplitude has a $t^{-1}$
factor since each contains a $\spa{l_i}.{l_{i+1}}^{-1}$ factor and,
for example,
$$
\hspace{1cm}\eqalign{
\spa{l_0}.{l_1}\; =\;
{[r | \Slash{l}_0\Slash{l}_1 | r] \over \spb{r}.{l_0}\spb{l_1}.r }
&\;=\;{[r | ( t \bar{r} +t^{-1}{r} +a_0)(  t \bar{r} +t^{-1}{r} +a_0-K_1) | r] \over \spb{r}.{l_0}\spb{l_1}.r }
\cr
&\;=\;{[r |  t \bar{r}(a_0-K_1)  +t a_0 \bar{r} +a_0(a_0-K_1)  | r] \over \spb{r}.{l_0}\spb{l_1}.r }
\cr
&\;=\;t\; {[r |   \bar{r}(a_0-K_1) + a_0 \bar{r}   | r] \over \spb{r}.{l_0}\spb{l_1}.r }
\;=\;t\; {-[r |   \bar{r} K_1   | r] \over \spb{r}.{l_0}\spb{l_1}.r }
\, ,
\cr}
\equn
$$
using the orthogonality properties of $r$, $\bar{r}$, $a_0$ and
$K_1$. The $\spb{r}.{l_i}$ factors cancel overall as the product of
tree amplitudes has no spinor weight in $l_i$. Thus, for each
particle circulating in the loop, the integrand has a $t^{-3}$
factor and $\rho_0$ must be extracted by expanding around this
triple pole.

For the $\NeqOne$ coefficients we present, summing over the particle
types leads to cancellations. Relative to the case of a scalar in
the loop, the $\NeqOne$ multiplet has an overall factor. Denoting
the three negative helicity external legs by $m_i$, this factor is,
$$
{ \left( \spa{l_0}.{m_1}\spa{l_1}.{m_2}\spa{l_2}.{m_3}-\spa{l_2}.{m_1}\spa{l_0}.{m_2}\spa{l_1}.{m_3} \right)^2
\over
 \spa{l_0}.{m_1}\spa{l_1}.{m_2}\spa{l_2}.{m_3}\spa{l_2}.{m_1}\spa{l_0}.{m_2}\spa{l_1}.{m_3}
} \; .
\equn
$$
Cancellations in the numerator give this expression an overall factor of $t^2$ implying that
the full  $\NeqOne$ integrand diverges as $t^{-1}$.
$\rho_0$ can then be extracted by taking a derivative:
$$\eqalign{
\rho_0 \,\sim \, { d  \over dt }  \Big(  t (  A_1A_2A_3
) \Big)_{t\, \longrightarrow\, 0}\,.\cr}\equn\label{dbydt}
$$
For $\NeqFour$ the overall factor is that of $\NeqOne$ squared and
thus introduces a $t^4$ factor. For these amplitudes it is thus
trivial to see that $\rho_0=0$ and there are no three-mass triangles
present in the expansion. This argument easily extends to show that
there are no three-mass triangles present in $\NeqEight$
supergravity~\cite{NoTriangle}.

\section{Canonical Forms}
We can use the techniques of the previous section to derive
canonical forms for evaluating the coefficients of three mass
triangles from the triple cut. In general we wish to expand the
product of tree amplitudes in the triple cut as a sum of standard
forms.
Let us take as the starting point a term of the form,
$$
{ \spa{b}.{\ell_0} \over \spa{a}.{\ell_0} }\, , \equn
$$
and let us carry out the parameterisation of  $\ell_0$ given in
(\ref{ParaEqn}) including a factor of $t^{-1}$ from the
measure to obtain the following integrand,
$$
{ \big(  t\spa{K_1}.{b} +\alpha_{01} \spa{K_2}.{b} \big)
\over
t \big(  t\spa{K_1}.{a} +\alpha_{01} \spa{K_2}.{a} \big)
}
=
 {1 \over t} { \spa{K_2}.{b}
\over \spa{K_2}.{a} } + { \big(  \spa{K_2}.{a}\spa{K_1}.{b}
-\spa{K_1}.{a}\spa{K_2}.{b} \big) \over
 \spa{K_2}.{a} \big(  t\spa{K_1}.{a} +\alpha_{01} \spa{K_2}.{a} \big)
 }\,,
\equn
$$
provided that $\spa{a}.{K_i} \neq 0$.
The contribution to the three-mass triangle is the residue at $t=0$
minus half the residue at $t\neq0$, namely,
$$
\eqalign{
& { \spa{K_2}.{b}
\over \spa{K_2}.{a} }
%%%
%%%\text{E: \it the formula does not seem to work unless this minus is a plus?}
%%%
-{ (  \spa{K_2}.{a}\spa{K_1}.{b} -\spa{K_1}.{a}\spa{K_2}.{b} )
\over 2
 \spa{K_2}.{a} \spa{K_1}.{a} }
\cr
&\hspace{1.2cm}={ (  \spa{K_2}.{a}\spa{K_1}.{b} +\spa{K_1}.{a}\spa{K_2}.{b} )
\over 2
 \spa{K_2}.{a} \spa{K_1}.{a} }
 \cr
&\hspace{1.2cm}={  \la a |( \hat K_1\hat K_2 -\hat K_2 \hat K_1)|b\ra
\over 2
\la a |\hat K_1 \hat K_2 | a\ra  }
={  \la a |( K_1 K_2 - K_2 K_1)|b\ra
\over 2
\la a | K_1  K_2 | a\ra  }
\; ,\cr}
\equn
$$
with the {$\hat K_i $ as defined as in eq.~(4.12).
When $\spa{a}.{K_1}=0$ (i.e. $\lambda_a \sim \lambda_{K_1}$) 
there is no $t\neq0$ pole and we have,
$$
{ \spa{\ell_0}.{b}
\over  \spa{\ell_0}.{K_1}  } \longrightarrow 
{ \spa{K_2}.{b}
\over  \spa{K_2}.{K_1} }
\; . 
\equn\label{NullFormEq}
$$

Next, consider expressions of the form,
$$
{ \spa{a}.{\ell_0} \spa{b}.{\ell_0} \over \la \ell_0 | K_1 K_2 | \ell_0 \ra }\,,
\equn
$$
which are evaluated  by replacing the $K_i$ by $\hat K_i$,
$$
\eqalign{
{ \spa{a}.{\ell_0} \spa{b}.{\ell_0} \over \la \ell_0 | K_1 K_2 | \ell_0 \ra }
&= { 1 \over (1-K_1^2K_2^2/\gamma^2)}
{ \spa{a}.{\ell_0} \spa{b}.{\ell_0} \over \la \ell_0 | \hat K_1 \hat K_2 | \ell_0 \ra }
={ 1 \over (1-K_1^2K_2^2/\gamma^2)}
{ \spa{a}.{\ell_0} \spa{b}.{\ell_0} \over \la \ell_0   K_1 \ra \spb{K_1}.{K_2} \la K_2  \ell_0 \ra }\,,
\cr
&=
{ \spa{a}.{\ell_0} \over \spb{K_1}.{K_2} (1-K_1^2K_2^2/\gamma^2)}
\Bigl( 
{  \spa{b}.{K_1} \over \la \ell_0   K_1 \ra  \la K_2  K_1\ra }+
{  \spa{b}.{K_2} \over \la K_2   K_1 \ra  \la K_2  \ell_0 \ra }
\Bigr)
\,,
\cr}
\equn
$$
%%%%%
which is two terms of the form~(\ref{NullFormEq}).
After some algebra, we can combine these terms to obtain,
$$
{ \spa{\ell_0}.a \spa{\ell_0}.b \over \la \ell_0 | K_1 K_2 | \ell_0 \ra }
\longrightarrow_{triangle}
  {  \la a | ( K_1 K_2 -K_2 K_1 ) | b \ra  \over  \Delta_3 }
 = {  \la a | [ K_1, K_2]  | b \ra  \over  \Delta_3 } \; .
\equn
\label{Canon3}
$$
We can extend this to,
$$
\eqalign{
{ \spa{\ell_0}.a \spa{\ell_0}.b \spa{\ell_0}.c \over \la \ell_0 | K_1 K_2 | \ell_0 \ra \spa{\ell_0}.d }
\longrightarrow_{triangle} &
 {  \la b |[ K_1, K_2]   | d \ra \la c |[ K_1, K_2]  | a \ra -\Delta_3 \spa{b}.d\spa{c}.a
  \over  2\Delta_3 \la d | K_1 K_2  | d \ra }
\cr
& + { \spa{d}.b \spa{d}.c   \la a | [ K_1, K_2]  | d \ra
\over 2 \la d | K_1 K_2  | d \ra^2  }\,.
\cr}
\equn\label{Canon4}
$$
This result will be sufficient to obtain the three-mass triangle coefficients for the $n$-point NMHV ${\cal N}=1$ contribution.

For a general ${\cal N}=1$ amplitude we also need,
$$
\eqalign{
{ [A|\ell_0|b\ra \spa{\ell_0}.c \over \spa{\ell_0}.d  }
\longrightarrow_{triangle}
&-{ [ A|a_0|d\ra  \big( \la d | [K_1,K_2]|b\ra \la d | [K_1,K_2]|c\ra  -\Delta_3\spa{d}.b\spa{d}.c \big)
\over  8 \la d |  K_1  K_2 | d \ra^2  }
\cr
&+{ [ A|a_0|b\ra \la d | [K_1,K_2]|c\ra +
 [ A|a_0|c\ra \la d | [K_1,K_2]|b\ra
 \over
4 \la d |  K_1  K_2 | d \ra }\,.
\cr}
\equn\label{Canon5}
$$

\section{$n$-Point NMHV  ${\cal N}=1$ Results}

We consider an $n$-point amplitude with three negative helicity
legs $m_i$. The triple cut vanishes unless there is precisely one
external negative helicity leg at each corner.
The product of the tree amplitudes will be,
$$
\eqalign{
&\sum_{h=0,\pm1/2}
A(\ell_0^h, r+1^+,\cdots , m_1^- , \cdots , s^+, -\ell_1^{-h} )
\times
A(\ell_1^h, s+1^+, \cdots , m_2^- , \cdots , t^+, -\ell_2^{-h} )\cr
& \hskip 7.5 truecm
\times
A(\ell_2^h, t+1^+, \cdots , m_3^- , \cdots , r^+ ,-\ell_0^{-h} )\cr
&\cr&\hspace{1cm}=
A(\ell_0^0, r+1^+, \cdots , m_1^- , \cdots , s^+,-\ell_1^{0} )
\times
A(\ell_1^0, s+1^+, \cdots , m_2^- , \cdots , t^+,-\ell_2^{0} )
\cr
& \hskip 4.5 truecm
\times
A(\ell_2^0,  t+1^+,\cdots , m_3^- , \cdots , r^+, -\ell_0^{0} ) \times \rho\,.
\cr}\equn
$$

The $\rho$-factor arises from summing over the multiplet and is,
$$
\eqalign{
\rho &
= {
\big(\spa{m_1}.{\ell_0} \spa{m_2}.{\ell_1} \spa{m_3}.{\ell_2}
-
\spa{m_1}.{\ell_1} \spa{m_2}.{\ell_2} \spa{m_3}.{\ell_0}  \big)^2
\over
\spa{m_1}.{\ell_0} \spa{m_2}.{\ell_1} \spa{m_3}.{\ell_2}
\spa{m_1}.{\ell_1} \spa{m_2}.{\ell_2} \spa{m_3}.{\ell_0}
}
\cr
& =
{ \spa{X}.{\ell_0}^2
\over
\spa{m_1}.{\ell_0} \spa{m_2}.{\ell_1} \spa{m_3}.{\ell_2}
\spa{m_1}.{\ell_1} \spa{m_2}.{\ell_2} \spa{m_3}.{\ell_0}  \spb{\ell_1}.{\ell_2}^2
}\,,
\cr}
\equn
$$
where,
$$
|X\ra = 
|m_1\ra \bigl( \la m_3 | K_3K_1|m_2\ra +\spa{m_2}.{m_3}(K_2^2-K_1^2)\bigr) 
+|m_3\ra  \la m_1 | K_1K_3|m_2\ra\,.
\equn
$$
The cut is of the form,
$$\eqalign{
{ 1 \over \prod_{i \neq r,s,t} \spa{i}.{i+1} } 
\times
{ \spa{m_1}.{\ell_0}\spa{m_1}.{\ell_1}
\over \spa{\ell_0}.{r+1} \spa{s}.{\ell_1}\spa{\ell_1}.{\ell_0} }
\times &
{ \spa{m_2}.{\ell_1}\spa{m_2}.{\ell_2}
\over \spa{\ell_1}.{s+1} \spa{t}.{\ell_2}\spa{\ell_2}.{\ell_1} }
\cr &
\times
{ \spa{m_3}.{\ell_2}\spa{m_3}.{\ell_0}
\over \spa{\ell_2}.{t+1}\spa{r}.{\ell_0}\spa{\ell_0}.{\ell_2} }
\times {\spa{X}.{\ell_0}^2 \over \spb{\ell_1}.{\ell_2}^2 }\,.
\cr}
\equn
\label{NMHVdiffls}
$$
In this we can combine,
$\spa{\ell_1}.{\ell_0}\spb{\ell_1}.{\ell_2}\spa{\ell_0}.{\ell_2}
=-\la\ell_0 | K_1 K_2 | \ell_0\ra $ and 
$\spa{\ell_2}.{\ell_1}\spb{\ell_1}.{\ell_2}=-K_3^2$. 
We can write (\ref{NMHVdiffls}) in terms of just one of the cut momenta using
identities of the form, 
$$
{ \spa{\ell_1}.b \over \spa{\ell_1}.a}
=
{ \spa{\ell_0}.{\ell_2} \spb{\ell_2}.{\ell_1} \spa{\ell_1}.b \over
\spa{\ell_0}.{\ell_2} \spb{\ell_2}.{\ell_1}\spa{\ell_1}.a}
= { \la \ell_0 | K_2\, K_3 | b \ra \over \la \ell_0 | K_2 \, K_3 | a \ra }
\equiv{\spa{\ell_0}.{b^{32}}\over \spa{\ell_0}.{a^{32}}},
$$
where we use the compact notation $|a^{ij}\ra \equiv K_jK_i|a\ra$.
This gives,
$$
\eqalign{
{ 1 \over K_3^2 \,\prod_{i \neq r,s,t} \spa{i}.{i+1} }
\times
{ \spa{m_1}.{\ell_0}\spa{m^{32}_1}.{\ell_0}
  \spa{m^{32}_2}.{\ell_0} \spa{m^{31}_2}.{\ell_0} 
   \spa{m_3}.{\ell_0}\spa{m^{31}_3}.{\ell_0}
\over \prod_{y\in Y_6} \spa{\ell_0}.{y} }
\times {\spa{X}.{\ell_0}^2 \over \la \ell_0 | K_1K_2 | \ell_0 \ra }\,,
\cr}
\equn
$$
where $Y_6=\{ r,r+1,s^{32},(s+1)^{32},t^{31},(t+1)^{31}\}$.
Using partial fractions this can be written as,
$$
{1 \over K_3^2 \,\prod_{i \neq r,s,t} \spa{i}.{i+1} }
 \times
\sum_{y \in Y_6}
{ \spa{m^{32}_1}.{y}
  \spa{m^{32}_2}.{y} \spa{m^{31}_2}.{y} 
   \spa{m_3}.{y}\spa{m^{31}_3}.{y}
\over \prod_{z\in Y_6,z\neq y } \spa{z}.{y} }
\times { \spa{m_1}.{\ell_0} \spa{X}.{\ell_0}^2
\over \spa{y}.{\ell_0} \la \ell_0 | K_1K_2 | \ell_0 \ra }\,.
\equn
$$

This is simply a sum of canonical forms~(\ref{Canon4}) and so the three-mass triangle 
coefficient is,
$$\hspace{-0.5cm}
\eqalign{
{ -i \over K_3^2 \prod_{i\neq r,s,t} {\spa{i}.{i+1} }  }
\sum_{y \in Y_6}&
{ \spa{m^{32}_1}.{y}
  \spa{m^{32}_2}.{y} \spa{m^{31}_2}.{y} 
   \spa{m_3}.{y}\spa{m^{31}_3}.{y}
\over \prod_{z\in Y_6,z\neq y } \spa{z}.{y} }
\times \cr & 
\biggl(
 {  \la m_1 | [ K_1 , K_2 ] | y\ra \la X | [K_1 , K_2 ]| X \ra
  \over  2\Delta_3 \la y | K_1 K_2  | y \ra }
 +  { \spa{y}.{X} \spa{y}.{m_1}   \la X | [ K_1 , K_2 ] | y \ra
\over 2\la y | K_1 K_2  | y \ra^2  }
  \biggr)\,.
\cr}
\equn
$$

\section{Beyond  NMHV  and ${\cal N}=0$}
We can, in principle, use the methods described above to obtain the three-mass
triangle coefficients for amplitudes beyond NMHV
or with less supersymmetry, i.e. ${\cal N}=0$.
In this section we outline how this may be performed.

In general the product of the tree amplitudes will be a sum of terms
which we treat individually.
The first step is to turn each term into a function of a single loop momentum,
say $\ell_0\equiv \ell$. Furthermore we will make this a function
depending predominantly on terms such as $\spa{a}.{\ell}$ rather than $\spb{a}.{\ell}$.
We can do this via replacements such as,
$$
\eqalign{
{ \spa{\ell_1}.b \over \spa{\ell_1}.a}
&=
{ \spa{\ell_0}.{\ell_2} \spb{\ell_2}.{\ell_1} \spa{\ell_1}.b \over
\spa{\ell_0}.{\ell_2} \spb{\ell_2}.{\ell_1}\spa{\ell_1}.a}
= { \la \ell | K_2\, K_3 | b \ra \over \la \ell | K_2 \, K_3 | a \ra }\,,
\cr
{\spb{\ell_1}.b \over \spb{\ell_1}.a }
&= { \la \ell | K_1 | b ] \over\la \ell | K_1 | a ] }\,,
\cr
{ \spb{\ell_0}.b \over \spb{\ell_0}.a }
& = { \spa{\ell_0}.{\ell_2}\spb{\ell_2}.{\ell_1}\spa{\ell_1}.{\ell_0} \spb{\ell_0}.b
\over \spa{\ell_0}.{\ell_2}\spb{\ell_2}.{\ell_1}\spa{\ell_1}.{\ell_0} \spb{\ell_0}.a }
={  \la \ell | K_2\,K_3\, K_1|b] \over  \la \ell | K_2\,K_3\, K_1|a] }\,.
\cr}
\equn
$$
While it is always possible to make the above replacements for the triple-cut, analogous replacements are not always possible for the two-particle cut.
Carrying out all possible replacement as described above,  each term in the cut can be written in the form,
$$
{ \prod^{n}_{i=1} \spa{A_i}.{\ell}
\over
 \prod^{n}_{j=1} \spa{B_j}.{\ell} \prod_{l=1}^p (\ell +Q_{l})^2 }
 \times  \prod^q_{k=1} [C_k|\ell|D_k\ra\,.
\equn
$$
We can tackle the massive propagators by utilising the identity,
$$
{ 1 \over (\ell+Q)^2  } \spb{C}.{\ell}
= { 1 \over (\ell+Q)^2  } { [C|K_1\,(K_1+Q)\,Q|\ell\ra \over \la \ell | K_1Q | \ell \ra }
-{ [C|K_1|\ell\ra \over \la \ell | K_1 Q | \ell \ra }\,.
\equn
$$
Using the parameterization (\ref{ParaEqn}) we see that the first and third terms are ${\cal{O}}(t^0)$ near $t=0$,
while the second is ${\cal{O}}(t^1)$ 
\footnote{It is worth noting that this counting would not have held had the numerators contained
$\la \ell | K_1 K_2 | \ell \ra$ rather than of $\la \ell | K_1 Q | \ell \ra$. Recalling that $\hat K_1$ and $\hat K_2$ are 
null vectors in the $(K_1,K_2)$ plane, we see that in this case there are cancellations within the
denominator which give it an extra overall factor of $t$.}.
 Multiple application of this identity leads to a sum of terms of the form, 
$$
{ \prod^{n+2p}_{i=1} \spa{A_i}.\ell \over \prod^{n+2p}_{j=1} \spa{B_j}.\ell } \left( \prod^{q-p}_{k=1} [C_k|\ell|D_k\ra \right)\,,
\equn
\label{masslessprod}
$$
together with terms that vanish at $t=0$ - these only contribute to box coefficients and can be neglected. 

In general the Yang-Mills amplitudes at each corner contribute an
effective overall momentum power of $\ell^1$. Thus the ${\cal N}=0$ amplitude contains terms
with momentum power up to $\ell^3$.  For ${\cal N}=1$ contributions, summing over the
multiplet cancels the two leading powers and we expect the three-point
integrals to go as $\ell^1$ and thus expect $q=p+1$.  In this case
(\ref{masslessprod}) is expressed as a sum of canonical terms
of the form evaluated in (\ref{Canon3}-\ref{Canon5}). For general ${\cal N}=0$
contributions we expect terms with $q=p+3$ which could be obtained using
higher power analogues of (\ref{Canon5}).

\section{Conclusions}

We have discussed a range of techniques that utilise the analytic
properties of one-loop amplitudes to generate the coefficients of
one-loop integral functions.  By combining these carefully we have generated
explicit expressions for the coefficients of the
three-mass triangle functions in any NMHV $n$-point $\NeqOne$ amplitude.

\section{Acknowledgments}

We wish to thank Zvi Bern, Lance Dixon and especially Harald Ita for many helpful conversations. This
research was supported in part by the Science and Technology
Facilities Council of the UK and in part by grant DE-FG02-90ER40542
of the US Department of Energy. NEJBB would like to thank the Niels
Bohr Institute and Swansea University for hospitality during the
completion of this work.

\end{document}